\documentclass[twocolumn,showpacs,preprintnumbers,amsmath,amssymb,prl]{revtex4}
\usepackage[english]{babel}
\usepackage{graphicx}% Include figure files
\usepackage{dcolumn}% Align table columns on decimal point
\usepackage{bm}% bold math

\begin{document}

\title{Excitation transfer and luminescence in porphyrin-carbon nanotube complexes}

% Force line breaks with \\
\author{G. Magadur$^{1,2}$, J.S. Lauret$^{1}$\footnote{email:lauret@lpqm.ens-cachan.fr, tel: +
33 1 47 40 55 99 } , V. Alain-Rizzo $^{2}$, C. Voisin$^{3}$, Ph.
Roussignol$^{3}$,\\ E. Deleporte$^{1}$ and J.A. Delaire$^{2}$}

\affiliation{$^{1}$Laboratoire de Photonique Quantique et
Mol\'eculaire, Institut d'Alembert, ENS Cachan, 61 avenue du
Pr\'esident Wilson 94235 Cachan Cedex,
France.\\
$^{2}$ Laboratoire de Photophysique et Photochimie
Supramol\'eculaires et Macromol\'eculaires,Institut d'Alembert,
ENS Cachan, 61 avenue du
Pr\'esident Wilson 94235 Cachan Cedex, France. \\
$^{3}$Ecole Normale Sup\'erieure, Laboratoire Pierre Aigrain, 24
rue Lhomond, 75005 Paris, France.}

\begin{abstract}
Functionalization of carbon nanotubes with hydrosoluble porphyrins
(TPPS) is achieved by "$\pi$-stacking". The porphyrin/nanotube
interaction is studied by means of optical absorption,
photoluminescence and photoluminescence excitation spectroscopies.
The main absorption line of the porphyrins adsorbed on nanotubes
exhibits a 120 meV red shift, which we ascribe to a flattening of
the molecule in order to optimize $\pi-\pi$ interactions. The
porphyrin-nanotube complex shows a strong quenching of the TPPS
emission while the photoluminescence intensity of the nanotubes is
enhanced when the excitation laser is in resonance with the
porphyrin absorption band. This reveals an efficient excitation
transfer from the TPPS to the carbon nanotube.

\end{abstract}
\maketitle

%%%%%%%%%%%%%%%%%%%%%%%%%%%%%%%%%%%%%%%%%INTRO%%%%%%%%%%%%%%%%%%%%%%%%%%%%%%%%%%%%%%%%%%%%%%%%%%%%%

Tailoring the properties of carbon nanotubes by functionalizing
their side-wall is one of the key issues towards the realization
of carbon nanotube-based optoelectronic or electronic devices.
Interactions between nanotubes and porphyrin molecules have
attracted much attention for about five years in connection with
applications such as nanotubes debundling \cite{rahman06},
nanotube sorting \cite{LI04}, photovoltaic cells \cite{hasobe06,
alvaro06, ehli06, herranz06} or biology \cite{guldi05}. Most of
the studies deal with either covalent \cite{alvaro06, herranz06}
or "$\pi$ - stacking" \cite{murakami03, rahman06, hasobe05,
hasobe06, LI04} functionalization. Changes in the porphyrin
optical absorption and emission spectra in the presence of
nanotubes have been reported \cite{rahman06,alvaro06, hasobe05,
hasobe06} while other studies do not report any changes
\cite{LI04, murakami03}. Several explanations have been proposed
for those changes including porphyrin protonation and H or
J-aggregates formation induced by  the nanotubes. In contrast and
despite the great interest of the question, little work has been
devoted to the alteration of nanotubes properties binded to
porphyrin molecules.

In this paper, we report on "$\pi$-stacking" functionalization of
nanotubes with hydrosoluble porphyrins. The interaction between
the porphyrin molecules and the nanotubes is studied by optical
absorption, photoluminescence and photoluminescence excitation
 spectroscopies. The Soret band of porphyrins
adsorbed on nanotubes exhibits a 120 meV red shift which is
interpreted as a consequence of the flattening of the molecule in
order to optimize $\pi-\pi$ interactions. A total quenching of the
porphyrin fluorescence is observed in the presence of nanotubes.
In the same time, the nanotubes photoluminescence is preserved
and enhanced when the excitation is tuned in resonance with the
Soret band of the "$\pi$-stacked" porphyrins. This result clearly
reveals an efficient excitation transfer between the porphyrin
molecules and the carbon nanotubes which is an avenue to many
applications.
\begin{figure}
    \centering
    \includegraphics[width=8.5 cm, height= 8cm]{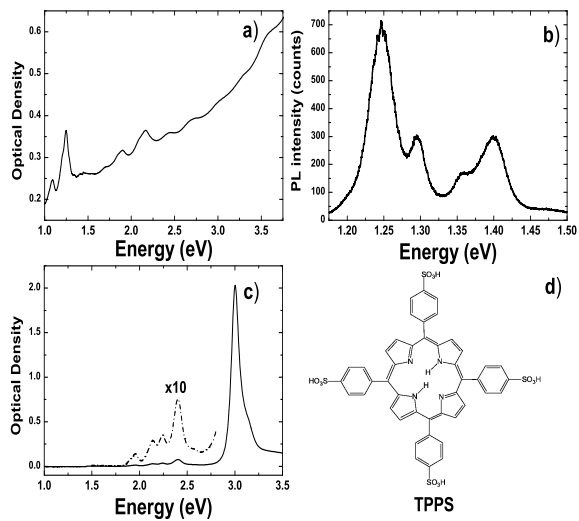}
  \caption{a) Optical absorption of CoMoCat nanotubes embedded in Sodium Dodecyl Sulfate micelles in a pH 8 Normadose buffer. b) Corresponding photoluminescence spectrum of
  CoMoCat nanotubes embedded in micelles (excitation energy: 2.331 eV). c) Optical absorption of TPPS in a pH 8 Normadose buffer(dashed curve: Q bands contribution rescaled
for clarity). d) Chemical structure of TPPS. The phenyl groups are
not in the plane of the macrocycle because of the steric
hindrance.}
    \label{fig 1}
\end{figure}

%%%%%%%%%%%%%%%%%%%%%%%%%%%%%%%%%%%%%%%%%Partie Exp?rimentale%%%%%%%%%%%%%%%%%%%%%%%%%%%%%%%%%%%%%%%%%%%%%%%%%%%%%
\begin{figure}
    \centering
\includegraphics[scale=.8]{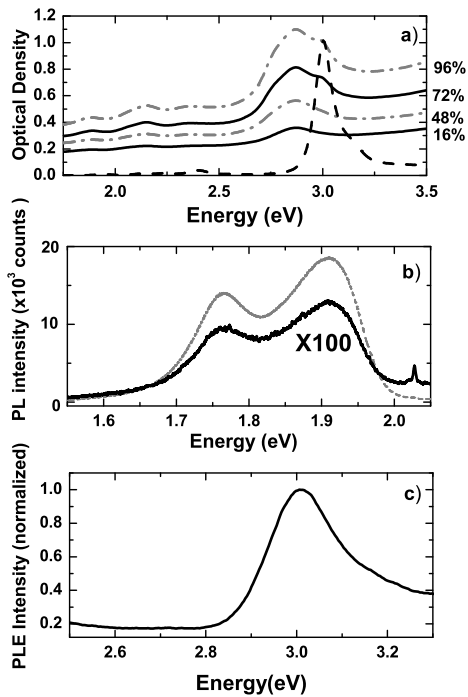}
\caption{a) Optical absorption of TPPS/CoMoCat nanotubes
suspensions around the Soret band energy for porphyrin
concentrations ranging from 16\% to 96\%. Free porphyrin (4\%)
absorption spectrum (black dashed curve) rescaled for clarity). b)
Photoluminescence spectra of TPPS (4\% \cite{concentration})
alone (grey) and in the presence of nanotubes (black). c)
Photoluminescence excitation spectrum of a porphyrin/nanotube
suspension detected at 1.77 eV.}
    \label{fig 2}
 \end{figure}

In this study we use CoMoCat carbon nanotubes which are
single-wall nanotubes (SWNT) with a mean diameter of about 0.8~nm
\cite{resasco1}. The nanotubes are functionalized by using the
5,10,15,20-tetrakis(4-sulfonatophenyl) porphyrin (TPPS). It was
synthesized according to conventional procedures described in the
literature \cite{Adler67}. It is soluble in a pH~8 Normadose
buffer (10$^{-2}$M) and its chemical structure is depicted in
figure~1d). Nanotubes functionalization is achieved by adding a
porphyrin solution (1mg/mL) in a suspension of SWNT
(0.1~mg.mL$^{-1}$) in a pH~8 Normadose buffer. The porphyrin
weight fraction is adjusted and increased from 4\%
(porphyrin/SWNT w/w) up to 96\%. After porphyrin addition, the
solution is sonicated for 3h with a dismembrator. The sample is
placed in a thermostat and cooled in water at 1$^{o}$C during
sonication.

Optical absorption spectra (OAS) are recorded with a
spectrophotometer (lambda 900 Perkin-Elmer). Two cw-laser diodes
at 3.062 eV (405 nm) and 2.331 eV (532 nm) are used as excitation
sources for photoluminescence (PL) experiments. The PL signal is
dispersed in a spectrograph (Spectrapro 2500i, Roper Scientific)
and detected by a Si CCD camera (Pixis 100B, PI Acton). When
detecting in the visible range, photoluminescence excitation
experiments (PLE) are performed with a spectrofluorometer
(Jobin-Yvon Fluoromax-3), while for detection in the near infrared
range a tunable frequency doubled Ti:Saphire laser is used as the
excitation source (2.75 eV - 3.1 eV).

%%%%%%%%%%%%%%%%%%%%%%%%%%%%%%%%%%%%%%%%%R?sultats%%%%%%%%%%%%%%%%%%%%%%%%%%%%%%%%%%%%%%%%%%%%%%%%%%%%%

The optical absorption spectrum of CoMoCat nanotubes embedded in
Sodium Dodecyl Sulfate (SDS) micelles at pH 8 is displayed in
figure~1a). The corresponding PL spectrum excited at 2.331~eV is
shown in figure~1b). Four main emission bands are observed at
1.24~eV, 1.29~eV, 1.36~eV and 1.40~eV, which are commonly
attributed to the (6,5), (8,3), (9,1) and (6,4) nanotubes
respectively \cite{resasco1,bachilo02}. The porphyrin optical
absorption spectrum is displayed in figure~1c). One can observe
the weak Q absorption bands (in the 1.7-2.5~eV range) and the
prominent Soret band (at 3~eV). The PL spectrum (grey curve in
figure~2b) exhibits two bands at 1.91~eV and 1.77~eV
corresponding to singlet-singlet transitions. Note that, once
rescaled, the optical absorption spectra of TPPS are identical for
concentrations between 4\% and 96\% \cite{concentration}, showing
that porphyrins do not aggregate up to this concentration.

The absorption spectrum of porphyrin/nanotube suspensions is shown
in figure~2a) for several porphyrin concentrations. The dashed
curve represents the free porphyrin (4\%) absorption spectrum
rescaled for clarity. For porphyrin concentrations below 72\%, the
Soret band peaks at 2.88~eV which corresponds to a red shift of
120~meV with respect to the absorption band of free porphyrins.
Since we have checked that no aggregation of the porphyrin occurs
for these concentrations, we assign this red shift to the
interaction between the porphyrin molecules and the nanotubes.

For a concentration of 72\% and up, a shoulder appears at 3~eV,
which corresponds to the Soret band of free porphyrins. This means
that nanotubes no longer bind to porphyrins above this
concentration and that additional porphyrins remain free in the
suspension. Further experiments are under progress to determine
more accurately the saturation threshold. Note that we cannot
observe the Q~bands of porphyrin/nanotube complexes since the
absorption is mainly due to the nanotubes in this energy range
(see figures~1a) and 1c)).

Figure~2b) shows the photoluminescence spectra (excited at
3.06~eV) of free porphyrins (4\%, grey line) and
porphyrin/nanotube complexes (4\%, black curve). The latter shows
a quenching of 99\% of the porphyrin line (around 1.8~eV)
indicating a strong interaction between "$\pi$-stacked"
porphyrins and nanotubes. The photoluminescence excitation
spectrum of a porphyrin/nanotube suspension (detection at 1.77~eV)
is shown in figure~2c). It peaks at 3~eV, that is at the energy
of the Soret band of the free porphyrin. Moreover, after
rescaling this spectrum is exactly the same as the one of free
porphyrin (not shown here). We conclude that the weak remaining
photoluminescence signal of the porphyrins actually arises from
residual free porphyrins.  As a consequence we deduce that
"$\pi$-stacked" porphyrins no longer emit light, supporting the
assumption of a strong nanotube-porphyrin interaction.

Several mechanisms may account for the red shift of the Soret
band. Protonation of TPPS leads to a red shift of about 140~meV
\cite{ohno93}. Since the experiments reported here are performed
in a pH~8 buffer and since the TPPS pKa are about 5, protonation
of the porphyrin can be ruled out. A red shift of the Soret band
may also stem from the formation of J-aggregates \cite{hasobe05,
hasobe06}, which is generally observed in acid environments and
leads to a red shift of about 500~meV. We tried to aggregate TPPS
in the pH~8 buffer by increasing gradually the porphyrin
concentration without any success. In contrast, in an acid medium
we were able to form J-aggregates with this protocol, in
agreement with previous studies \cite{ohno93}. Thus we deduce
that J-aggregates cannot exist in our experimental conditions and
therefore cannot account for the observed red shift.

A red shift of the Soret band has also been reported for
porphyrins laid on a substrate \cite{chernia99} or self-assembled
into nanoparticles \cite{ou07}. In that case the molecules
flatten to enhance the interaction with the substrate or with
other porphyrin molecules. The phenyl groups, which make a
90$^{o}$ angle with the plane of the macrocycle for free
porphyrins in solution, tend to tilt and to reduce this angle
when interacting with the substrate or other molecules. This
behavior leads to a red shift of the Soret band of about 200~meV
for a tilt of 30$^{\mathrm{o}}$ \cite{chernia99}. In our
experiments, we assign the observed red shift of the Soret band
to a flattening of the porphyrin in contact with a nanotube.

As a consequence, this red shift of the Soret band should depend
on the nanotube diameter~; in fact, the larger the nanotube
diameter, the more the porphyrin has to flatten in order to
optimize the $\pi$-$\pi$ interaction with the nanotube. Thus, the
red shift should increase with the tube diameter \cite{chernia99}.
Preliminary results show that TPPS adsorbed on nanotubes with a
mean diameter of 1.2~nm (synthesized by the electric arc method)
exhibit a 170~meV red shift of the Soret band. Compared to the
previously mentioned CoMoCat nanotubes, we observe a 40\%
increase of the red-shift for a 50\% increase of the nanotube
diameter, which is qualitatively consistent with our
interpretation.

Nevertheless, it has been shown that the photoluminescence signal
of porphyrins is preserved when they flatten to form nanoparticles
\cite{ou07}. Thus, the strong quenching of the porphyrin
fluorescence in the presence of nanotubes cannot be due to the
deformation of the molecules. This rather seems to indicate a
strong electronic interaction between porphyrins and nanotubes.

\begin{figure}
    \centering
\includegraphics[scale=.9]{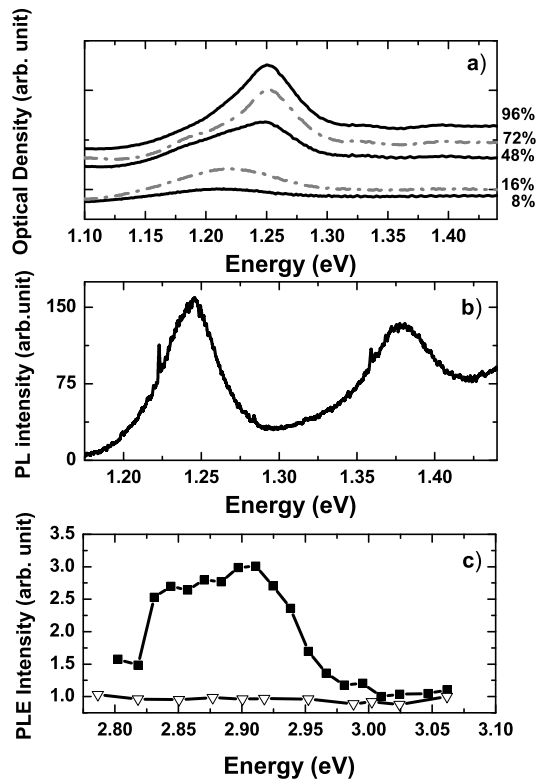}
\caption{a) Optical density of TPPS/CoMoCat nanotubes suspension
around first excitonic transition of nanotubes for different
concentrations of porphyrins (from 8\% to 96\%). Curves have been
arbitrary shifted on the Y-axis for clarity. b) Photoluminescence
signal of CoMoCat nanotubes in presence of TPPS excited at 2.331
eV. c) Photoluminescence excitation spectra of a
porphyrin/nanotube suspension (black squares) and of nanotubes
embedded in SDS micelles (open triangles) detected at 1.246 eV.}

    \label{fig 4}
 \end{figure}

Optical absorption and fluorescence were recorded similarly in
the spectral range of the nanotube first excitonic transitions
usually labelled "S$_{11}$". The evolution of the nanotube
absorption as a function of the porphyrins concentration is
depicted in figure~3a). We first observe that the specific
absorption band of the nanotubes is better resolved when the
porphyrin concentration is higher. In order to be more
quantitative, we define an aspect ratio "R" such as $R =
\frac{OD_{max}}{OD_{min}}$ where $OD_{max}$ is the optical
density at the maximum of the band and $OD_{min}$ is the optical
density at the high energy foot (1.32~eV) of the band. We get $R=
1.09$ for a porphyrin concentration of 8\% and $R = 1.37$ for a
concentration of 72\%. For comparison, for nanotubes embedded in
micelles (reference material, see figure~1a) ), we get $R = 1.43$
\cite{note1}.

The absorption band consists of a shoulder at 1.19~eV and a
maximum at 1.251~eV (for TPPS concentration up to 72\%). For
higher porphyrin concentrations, the shoulder disappears (see the
96\% curve in figure~3a)) while the absorption at 1.251~eV
increases. This seems to indicate that TPPS interacts
preferentially with some specific types of nanotubes (i.e.
chiralities).

Further insight can be gained by photoluminescence measurements.
Figure~3b) shows the photoluminescence signal of CoMoCat
nanotubes functionalized with TPPS. In contrast to the case of
covalent grafting \cite{herranz06}, the PL signal of the
nanotubes is preserved after functionalization. Transmission
electronic microscopy analysis shows that the suspensions contain
some individual nanotubes and some remaining bundles. If
porphyrins actually foster the debundling of nanotubes, one
expects an enhancement of the quantum efficiency for
functionalized nanotubes
\cite{oconnel02,lauret1,lauret2,berger07}. Indeed, we
qualitatively observe that the higher the porphyrin
concentration, the larger the nanotubes PL intensity.

The PL signal of functionalized nanotubes exhibits two bands at
1.246~eV and 1.389~eV. Compared to the PL signal of nanotubes
embedded in surfactant, two bands -corresponding to (8,3) and
(9,1) nanotubes- are missing in the spectrum of functionalized
nanotubes. This is consistent with our previous conclusion that a
partial nanotube sorting is achieved.

Moreover, the PLE spectrum of porphyrin/nanotube complexes
detected at 1.246 eV (see figure~3c)) shows that the PL signal of
the functionalized nanotubes is greatly enhanced when the sample
is excited at 2.88~eV. In contrast, the PLE spectrum of nanotubes
embedded in SDS micelles is flat in this spectral range (see
figure~3c)). We assign this PLE band at 2.88 eV to the Soret band
of the "$\pi$-stacked" porphyrins, as shown from the absorption
spectra (see figure 2a)). This clearly means that the PL signal
arises from functionalized nanotubes. Furthermore, it shows that
light absorbed on "$\pi$-stacked" porphyrin states results in a
population on excited states of nanotubes. Therefore, it brings
evidence for an efficient excitation transfer from the porphyrins
to the nanotubes within the ``$\pi$-stacked'' complexes.

In summary, we have demonstrated the functionalization of CoMoCat
nanotubes by TPPS porphyrins. Porphyrins "$\pi$-stacked" on
nanotubes exhibit a red shift of the Soret band, which has been
interpreted in terms of a flattening of the porphyrin molecule in
order to optimize the $\pi-\pi$ interactions with the nanotube.
Moreover, we have shown that TPPS interacts preferentially with
some specific classes of nanotubes. In contrast to covalent
functionalization, the photoluminescence signal of nanotubes is
preserved. In the same time, "$\pi$-stacked" porphyrins no longer
emit light, but the nanotube fluorescence is enhanced when the
excitation energy is tuned in resonance with the absorption band
of the porphyrins bringing evidence for a strong excitation
transfer from TPPS to the nanotubes. In view of potential
applications, the understanding and the optimization of this
excitation transfer would be valuable. Time-resolved experiments
are under progress to investigate further the electronic states
involved in this coupling, as well as its nature and its dynamics.

\textbf{Acknowledgments}:  The authors are grateful to I. Leray
for providing the porphyrin molecules and to D.E. Resasco for
providing the CoMoCat nanotubes. LPQM, PPSM and LPA are "Unit\'es
mixtes" de recherche associ\'ees au CNRS (UMR8537; UMR8531;
UMR8551). This work was supported by the GDR-E "nanotube"
(GDRE2756) and grant "C'Nano IdF Phototube" from "R\'egion Ile de
France".


\begin{thebibliography}{99}
\bibitem{concentration} In order to compare quantitatively
suspensions with and without nanotubes, we label free porphyrin
suspensions with the expression ``X\%''. It corresponds to the
porphyrin concentration that would be used in a X\%
porphyrin/nanotube suspension where X represents the ratio
porphyrin/SWNT w/w for an initial NT concentration of 0.1mg/mL.
\bibitem{note1}
The authors want to point out that, even if the absolute value of
"R" is subject to uncertainty, the general trend is reproducible
indicating that porphyrins help solubilizing nanotubes, even if
the solubilizing efficiency is not as high as for surfactants.
\end{thebibliography}
\end{document}